\documentclass[conference]{IEEEtran}
\IEEEoverridecommandlockouts
\usepackage{cite}
\usepackage{amsmath,amssymb,amsfonts}
\usepackage{algorithmic}
\usepackage{graphicx}
\usepackage{textcomp}
\usepackage{xcolor}
\usepackage{bm}
\def\BibTeX{{\rm B\kern-.05em{\sc i\kern-.025em b}\kern-.08em
    T\kern-.1667em\lower.7ex\hbox{E}\kern-.125emX}}
\begin{document}

\title{MIMO-OFDM Scheme design for Medium Voltage Underground Cables based Power Line Communication}

\author{\IEEEauthorblockN{Mengyuan Cheng, Kai Wan, Fan Wei, Fuyong Zheng, and Yongpeng Wu}

\thanks{M. Cheng, F. Wei, and Y. Wu are with the Department of Electrical Engineering, Shanghai Jiao Tong University, Minhang 200240, China (Email: my$\_$cheng@sjtu.edu.cn; weifan89@sjtu.edu.cn; yongpeng.wu@sjtu.edu.cn).}
\thanks{K. Wan is with ICT Department, Global Energy Interconnection Research Institute, Future Science City, Beijing 102209, China (Email: wankai@geiri.sgcc.com.cn).}

\thanks{F. Zheng is with Technology and ICT Department, Jiangxi Electric Power
Company, Nanchang 330029, China (Email: zhengfuyong@jx.sgcc.com.cn).}

\thanks{Y. Wu is the corresponding author of the paper.}

}\maketitle

\begin{abstract}
Power line communication (PLC) provides intelligent electrical functions such as power quality measurement, fault surveys, and remote control of electrical  network. However, most of research works have been done in low voltage (LV) scenario due to the fast development of in-home PLC. The aim of this paper is to design a MIMO-OFDM based transmission link under medium voltage (MV) underground power line channel and evaluate the performance. The MIMO channel is modeled as a modified multipath model in the presence of impulsive noise and background noise. Unlike most literatures on MIMO power line transmission, we adopt spatial multiplexing  instead of diversity to increase the transmission rate in this paper. The turbo coding method originally designed for LV power line communication is used in the proposed transmission system. By comparing the BER performance of MIMO-OFDM system with and without the turbo coding, we evaluate its applicability in MV power line communication. The effect of frequency band varying on the PLC system's performance is also investigated.
\end{abstract}

\begin{IEEEkeywords}
power line communication (PLC), MIMO, OFDM, multi-conductor, medium voltage underground cables, impulsive noise, spatial multiplexing, turbo coding
\end{IEEEkeywords}

\section{Introduction}
Since the international standard (IEEE 1901) was adopted in 2010\cite{b1}, power line served as a novel propagation medium, has become a promising candidate
for the information transmission. Although the existing power-supply network was not designed for communication purposes originally, power line communication (PLC) enjoys the advantage of low implementation cost since there is no need for extra installation or maintenance. Moreover, electricity networks spread over wider areas than typical wireless infrastructure. However, in PLC , the channels are normally contaminated by impulsive noises.

Orthogonal frequency division multiplexing (OFDM) is robust to inter-symbol interference (ISI) caused by frequency-selective channels,  and therefore is widely utilized in PLC. In \cite{b2}, the bit-error-rate (BER) performance of OFDM PLC considering impulsive noise is demonstrated and the results show that the OFDM-based PLC systems are subject to  much higher error rate in the presence of impulsive noise. In order to effectively combat the poor channel conditions, channel coding is indispensable due to its capability for error correction  and anti-jamming.

When combined with multiple-input multiple-output (MIMO), the coverage and capacity of the OFDM PLC transmissions could be enhanced significantly. MIMO-OFDM based PLC systems have been investigated extensively in recent research works\cite{b3,b4,b5,b6}, where space diversity, turbo coding, and low-density parity-check (LDPC) coding were evaluated in low voltage (LV) PLC scenario. However, to the best of the authors' knowledge, few technical effort considers coded MIMO-OFDM based medium voltage (MV) PLC system due to the difficulties of the channel measurement of MV cables.

3-phase 3-wire power lines are normally used for MV power line transmissions with two main configurations: MV overhead, and MV underground. In this paper, we focus mainly on underground MV cables to design a 3$\times$3 MIMO system. With the assistance of State Grid Corporation of China, we have access to the channel modeling parameters of a section of buried cable of 1 km long in Ganzhou city, Jiangxi province, China. For the channel model, we adopt the modified multipath model for broadband MIMO PLC which takes into consideration the  coupling effects between inner conductors\cite{b7}. In other words, there exists crosstalk among the MIMO channels. Based on the coupled channel model, we present a comprehensive communication link for MV power line communication, including turbo coding, spatial multiplexing, channel estimation, soft demodulation, and MMSE detection. Furthermore, the optimal frequency band is investigated by comparing the BER performances of different frequency bands within the frequency range of 0.5 MHz-6.5 MHz. We have found that the best performance is achieved on frequency band of 3.0 MHz-5.0 MHz, whereas 1.0 MHz-3.0 MHz comes second.

The remainder of this paper is organized as follows: MIMO channel modeling and impulsive noise are presented in Section \uppercase\expandafter{\romannumeral2}. Section \uppercase\expandafter{\romannumeral3} shows the turbo coded MIMO-OFDM system. Section \uppercase\expandafter{\romannumeral4} shows the simulation results. Concluding remarks are drawn in Section \uppercase\expandafter{\romannumeral5}.

\section{ MIMO Channel Modeling $\&$ Noise Characterization}
\subsection{MIMO Channels with Crosstalk}
In this paper, the MIMO channel is established for the selected section of MV underground cable whose type is 3-phase 3-wire multi-conductor. Fig. 1 illustrates the transversal section of the cable we used, which consists of an external shield, and 3 inner conductors for 3 phases respectively. The interstices between conductors are filled with polyvinyl chloride (PVC) dielectric material. The signal is injected between the shield and the core of one phase line with the shield normally connected to the ground at the end of the cable.

\begin{figure}[htbp]
\centerline{\includegraphics[width=4cm,height=3.3cm]{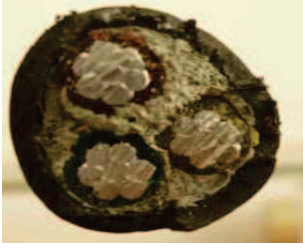}}
\caption{3-phase 3-wire multi-conductor power cable.}
\label{fig2}
\end{figure}

3$\times$3 MIMO system can be obtained by simply regarding the phase-A (green) cable as channel 1, phase-B (yellow) as channel 2, and phase-C (red) as channel 3. The modified multi-path MIMO channel model proposed in \cite{b7} considers the coupling effects among conductors on the basis of Zimmermann¡¯s model\cite{b13}, which is widely used in SISO channel. The channel transfer function (CTF) for the path from the $i$th transmit antenna to the $j$th receive antenna is given by \cite{b7}
\begin{equation}
H_{i,j}(f)=\sum_{p=1}^{N_p}{g_p}e^{-\alpha_{i,j}(f)d_p}e^{-j2\pi f\tau_p},
\end{equation}
where $i = 1,2,3$ and $j = 1,2,3$. Therefore, the overall transfer function matrix can be written by
\begin{equation}
\mathbf{H}_{(MIMO)}=
\begin{bmatrix}
   H_{1,1}(f) & H_{1,2}(f) & H_{1,3}(f) \\
   H_{2,1}(f) & H_{2,2}(f) & H_{2,3}(f) \\
   H_{3,1}(f) & H_{3,2}(f) & H_{3,3}(f)
\end{bmatrix},
\end{equation}
where the non-zero anti-diagonal terms $H_{i,j}$ (where $i\neq j$) indicate crosstalk between co-channels (indicated by  diagonal terms $H_{1,1}$, $H_{2,2}$ and $H_{3,3}$). The attenuation constant $\alpha_{i,j}$ in the attenuation portion of $H_{i,j}$ can be extracted from
\begin{equation}
\alpha_{i,j}={\rm Real} \left\lbrace \left( \sqrt{(\mathbf{R}''+j\omega \mathbf{L}'').\ast(\mathbf{G}''+j\omega \mathbf{C}'')}\right)_{i,j}\right\rbrace
\label{Eq3},
\end{equation}
where the element-wise operation $.\ast$ in \eqref{Eq3} is used to extract $\alpha_{i,j}$ from the transmission line matrices $\mathbf{R}''$, $\mathbf{L}''$, $\mathbf{C}''$, and $\mathbf{R}''$ which represent the mutual interactions between conductors \cite{b14}. The underground transmission line can be characterized by the equivalent per-unit-length (p.u.l) parameter model \cite{b14}. Parameters involved in modeling the 3$\times$3 MIMO are listed in Table I. All these parameters are measured on the selected MV underground power line cable.
\begin{table}[htbp]
\caption{Modeling Parameters of MV Cable}
\begin{center}
\begin{tabular}{|l|l|}
\hline
\textbf{Parameters} &  \textbf{Value}\\
\hline
\hline
Permeability of conducting material $\mu_c$ &  $4\times\pi\times10^{-7}$  H/m\\
\hline
Conductivity of conducting material $\sigma_c$ & $5.882\times 10^{7}$  S/m\\
\hline
Conductor spacing $D$ &  0.0197 m\\
\hline
Permeability of dielectric material $\mu_o$ & $4\times\pi\times10^{-7}$  H/m \\
\hline
Permittivity of dielectric material $\varepsilon_o$ & $2.2/(36\times\pi\times 10^{9})$  F/m\\
\hline
Conductor radius $r$ &  $0.003775$ m\\
\hline
Skin depth of the conducting material ${\rm tan}\delta $ &  $5.0\times 10^{-4}$\\
\hline
\end{tabular}
\label{tab1}
\end{center}
\end{table}

The resistance matrix is
\begin{equation}
R''=
\begin{bmatrix}
   r_1+r_0 & r_0 & r_0 \\
   r_0 & r_2+r_0 & r_0 \\
   r_0 & r_0 & r_3+r_0
\end{bmatrix},
\end{equation}
where $r_0$ is the ground resistance while $r_1$, $r_2$ and $r_3$ are the resistances for 3 phase inductors per unit, and are computed as $r_1=r_2=r_3=\frac{1}{2}\sqrt{\frac{\pi f\mu_c}{\sigma_c}}$.

The inductance matrix is
\begin{equation}
L''=
\begin{bmatrix}
   l_{11} & l_{12} & l_{13} \\
   l_{21} & l_{22} & l_{23} \\
   l_{31} & l_{32} & l_{33}
\end{bmatrix},
\end{equation}
where the diagonal terms indicate self-inductance and can be computed as:
$l_{11}=l_{22}=l_{33}=\frac{\mu_0}{2\pi}\ln \frac{D}{r}$. The anti-diagonal terms are the mutual inductances between conductors and are given as
\begin{equation}
\begin{aligned}
& l_{12}=l_{21}=k\sqrt{l_{11}l_{22}}\\
& l_{13}=l_{31}=k\sqrt{l_{11}l_{33}}\\
& l_{23}=l_{32}=k\sqrt{l_{22}l_{33}}
\end{aligned} \quad,
\end{equation}
where $k\in[0,1]$ is a constant representing the coefficient of coupling.

The capacitance matrix is
\begin{equation}
C''=
\begin{bmatrix}
   c_{11} & -c_{12} & -c_{13} \\
   -c_{21} & c_{22} & -c_{23} \\
   -c_{31} & -c_{32} & c_{33}
\end{bmatrix},
\end{equation}
where the anti-diagonal terms $c_{12}=c_{13}=\ldots=c_{23} $ (all referred to as $c_m$) represent the mutual capacitances and can be computed as: $c_m=4\pi\varepsilon_0$. The diagonal terms $c_{11}$, $c_{22}$ and $c_{33}$ can be obtained by
\begin{equation}
\begin{aligned}
& c_{11}=c_{1G}+c_{12}+c_{13}\\
& c_{22}=c_{2G}+c_{21}+c_{23}\\
& c_{33}=c_{3G}+c_{31}+c_{32}\\
\end{aligned}\quad,
\end{equation}
where $c_{NG}=\frac{2\pi\varepsilon_0}{\ln\frac{D}{r}}$ $(N=1,2,3)$.

The conductance matrix is
\begin{equation}
G''=
\begin{bmatrix}
   g_{11} & -g_{12} & -g_{13} \\
   -g_{21} & g_{22} & -g_{23} \\
   -g_{31} & -g_{32} & g_{33}
\end{bmatrix},
\end{equation}
where mutual conductance $g_{12}$, $g_{13}$, $\ldots$ $g_{23}$ (all denoted by $g_m$), are calculated by: $g_m=2\pi fc_m\mathrm{tan}\delta$. And the diagonal terms $g_{11}$, $g_{22}$ and $g_{33}$ can be computed as
\begin{equation}
\begin{aligned}
& g_{11}=g_{1G}+g_{12}+g_{13}\\
& g_{22}=g_{2G}+g_{21}+g_{23}\\
& g_{33}=g_{3G}+g_{31}+g_{32}\\
\end{aligned}\quad,
\end{equation}
where $g_{NG}=2\pi f c_{NG}\mathrm{tan}\delta$ $(N=1,2,3)$.

For further details of computing the channel coefficients,  please refer to \cite{b7}.

\subsection{Middleton Class-A Model}
Unlike the wireless communication, the noise in PLC cannot be simply described as additive white Gaussian noise (AWGN). Instead, it consists of the background and impulsive noise. The background noise can be regarded as Gaussian noise\cite{b9}. While for impulsive noise, Middletons Class A noise (AWCN) \cite{b10} based on the Poisson-Gaussian model suffices to be an accurate model. As discussed in\cite{b11}, the probability density function (PDF) of impulsive noise can be modeled as a Poisson weighted sum of Gaussian distributions:
\begin{equation}
p(n)=\sum_{m=0}^\infty\frac{A^m\cdot e^{-A}}{\sqrt{2\pi}\cdot m!\cdot \sigma_m}\cdot \mathrm{exp}(-{\frac{n^2}{2\sigma^2_m}}),
\end{equation}
where $m$ is the number of active interferences (or impulses), $A$ is the impulse index that indicates the average number of impulses during interference time. The bigger $A$ is, the more impulsive the noise gets. Conversely, the noise tends towards AWGN  when $A$ is small. $\sigma_m^2$is given by
\begin{equation}
\sigma_m^2=\sigma^2\cdot\frac{\frac{m}{A}+\Gamma}{1+\Gamma},
\end{equation}
where
\begin{equation}
\sigma^2=\sigma_g^2+\sigma_i^2   , \qquad      \Gamma=\frac{\sigma_g^2}{\sigma_i^2}\quad.
\end{equation}
$\sigma^2$ is the total noise power equal to the sum of the Gaussian noise power $\sigma_g^2$ and the impulsive noise power $\sigma_i^2$. $\Gamma$ is the Gaussian factor and the value is low when impulsive component prevails.

In \cite{b12}, the authors give the impulsive noise sample in the form of
\begin{equation}
n=x_g+\sqrt{K_m}\cdot w,
\end{equation}
where $x_g$ is the white Gaussian background noise sequence with zero mean and variance $\sigma_g^2$, $w$ is the white Gaussian sequence with zero mean and variance $\frac{\sigma_i^2}{A}$. $K_m$ is the Poisson distributed sequence, whose PDF is characterized by the impulsive index $A$.

\section{Turbo Coded MIMO-OFDM Scheme}
Fig. 2 shows the block diagram of a turbo coded MIMO-OFDM system with 3 transmit antennas (3 phases) and 3 receive antennas (3 phases). In the transmitter side, the input bit stream is fed into the QPSK modulator after being turbo encoded. The modulated symbols are further divided into 3 data streams through serial-to-parallel conversion. Subsequently, each data stream is separated into date blocks of K length. Every data block is processed by OFDM modulation (IFFT) and cyclic prefix (CP) insertion sequentially. The reverse process is carried out in the receiver side after data symbols go through the PLC MIMO channels. There are two extra blocks, i.e, channel estimation and detector, between OFDM demodulator (FFT) and parallel-to-serial conversion blocks. Training sequences based channel estimation is used in this paper, and the detector is consists of two components: detection and soft demodulation. Due to the presence of turbo coding in the system, soft demodulation is indispensable for the turbo decoding process.

\begin{figure}[htbp]
\centerline{\includegraphics[width=9cm,height=4cm]{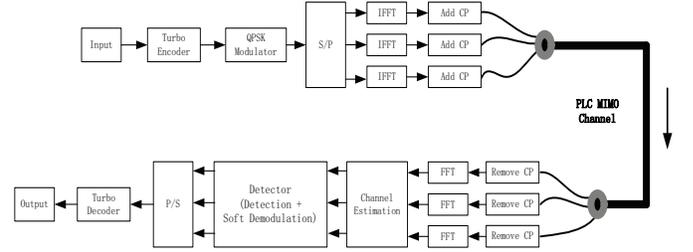}}
\caption{Block diagram of turbo coded MIMO-OFDM system}
\label{fig2}
\end{figure}

\subsection{Turbo Encoder Structure}
The turbo encoder is made up of two component encoders (ENC1, ENC2), which are cascaded in parallel through an interleaver. The structure adopted by \textit{low voltage power line communication interoperability technical specification}\cite{b8} is shown in Fig. 3. For systematic codes, the input of every two information bits $(u_1,u_2)$ will output the identical system bits $(u_1,u_2)$ and parity check bits $(p,q)$. The essential reasons for the excellent performance of turbo codes are summarized as follows: firstly, the interleaver used between the two component encoders equips the codeword with a large free distance; secondly, system bits are doubly constrained by the two identical encoders. Accordingly, the two component decoders iteratively decode information bits based on the maximum a posteriori probability (MAP) criterion by interchanging soft metric values between each other.
\begin{figure}[htbp]
\centerline{\includegraphics[width=6.5cm,height=4.3cm]{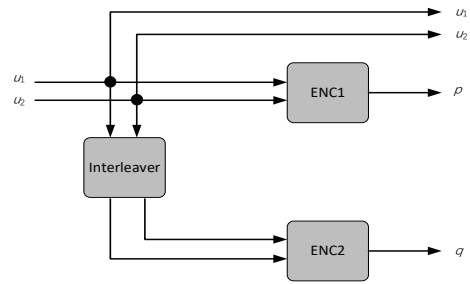}}
\caption{Encoding structure}
\label{fig3}
\end{figure}
\begin{figure}[htbp]
\centerline{\includegraphics[width=9cm,height=4cm]{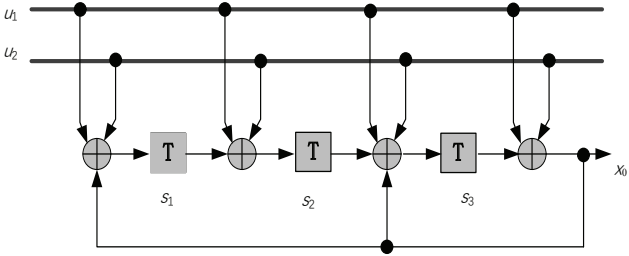}}
\caption{ENC1/ENC2 structure}
\label{fig4}
\end{figure}

As it is shown in Fig. 4, an 8-state encoder is utilized in each component encoder. Every two input bits are mapped to $(u_1,u_2)$ and output two parity check bits $(p,q)$. The specific algorithm of each component encoder is as follows\cite{b8}:
\begin{itemize}
\item Set the initial state of the registers $\mathbf{S}_0=[S_{01},S_{02},S_{03}]$ as $[0,0,0]$.
\item Fed bits to component encoder (ENC2's input bits are interleaved) until the last bit to get the final state $\mathbf{S}_N=[S_{N1},S_{N2},S_{N3}]$
\item Define the matrix $\mathbf{M}$. The size of physical layer (PHY) block ($\mathrm{PB}\_\mathrm{Size}$) is set to 264 in this paper, which makes
\begin{equation}
\mathbf{M}=
\begin{bmatrix}
1 & 0 & 1\\
1 & 1 & 1\\
1 & 1 & 0\\
\end{bmatrix}.
\end{equation}
Beyond that, $\mathbf{M}=\begin{bmatrix}
0 & 0 & 1\\
1 & 0 & 1\\
1 & 1 & 1\\
\end{bmatrix}$ for $\mathrm{PB}\_\mathrm{Size}$ of 16, 72, and 520. For $\mathrm{PB}\_\mathrm{Size}$ of 136, $\mathbf{M}=\begin{bmatrix}
0 & 1 & 1\\
1 & 0 & 0\\
0 & 1 & 0\\
\end{bmatrix}$.\\
Then, $\mathbf{S}_0'=\mathbf{S}_N\times \mathbf{M}$
\item Re-enter the input bits into the component encoder, whose initial state is calculated in the previous step. After that, parity bits can be obtained from the final state $\mathbf{S}_N'=\mathbf{S}_0'$
\end{itemize}

The input bits are interleaved by turbo interleaver and then fed to ENC2. The interleaving process is based on the unit of two bits, therefore the interleaving length is equal to half of the number of input bits. As listed in Table II, different interleaving length corresponds to different $\mathrm{PB}\_\mathrm{Size}$. For further detailed process of encoding and interleaving, please refer to \cite{b8}. Afterwards, the parity bits should be punched according to the code rate, which is either 1/2 or 16/18 in the used turbo coding method.
\begin{table}[htbp]
\caption{Simulation Parameter}
\begin{center}
\begin{tabular}{|c|c|}
\hline
$\mathrm{\mathbf{PB}}\_\mathrm{\mathbf{Size}}$ \textbf{(number of bytes)} &   \textbf{ Interleaving length}\\
\hline
\hline
16& 64\\
\hline
72  &  288\\
\hline
136  & 544 \\
\hline
264 & 1056\\
\hline
520  & 2080\\
\hline

\end{tabular}
\label{tab1}
\end{center}
\end{table}

\subsection{Soft Demodulation}
Soft demodulation is used to calculate log likelihood ratio (LLR) of each bit using the estimation value of transmitted signal $\hat{\bm{s}}$ and the variance $\bm{\sigma}^2$ of residual interference plus noise. $\hat{\bm{s}}$ and $\bm{\sigma}^2$ are obtained by minimum mean square error (MMSE) detection module performed before soft demodulation.

A transmitted signal $s_k$ of the $k^\mathrm{th}$ transmit antenna is assumed to be mapped from an $M_c\times 1$ vector $\mathbf{d}$ containing $M_c$ bits. The posterior LLR of each bit in $s_k$ is given by
\begin{equation}
\begin{aligned}
L_D(d_j\mid \hat{s_k})&=\ln\frac{p(\tilde{d}_j=+1\mid \hat{s_k})}{p(\tilde{d}_j=-1\mid \hat{s_k})}\\
&=\ln\frac{\sum\limits_{s_k=\alpha(\mathbf{d})\in S_{j,1}}p(\hat{s}_k\mid s_k)P[s_k=\alpha(\mathbf{d})]}{\sum\limits_{s_k=\alpha(\mathbf{d})\in S_{j,0}}p(\hat{s}_k\mid s_k)P[s_k=\alpha(\mathbf{d})]},
\end{aligned}
\label{Eq9}
\end{equation}
where, $S_{j,1}=\{ \mathbf{d}\mid d_j=1\}$, $S_{j,0}=\{ \mathbf{d}\mid d_j=0\}$; and $\tilde{d}_j$ is defined as
\begin{equation}
\tilde{d}_j\triangleq \left\lbrace
\begin{array}{lr}
+1,\quad d_j=1  &  \\
-1,\quad d_j=0\quad.
\end{array}
\right.
\end{equation}

With no a priori information for soft demodulation, $P[s_k=\alpha(\mathbf{d})]=1/M_c$ in \eqref{Eq9}. The likelihood function $p(\hat{s}_k\mid s_k)$ can be calculated by the variance $\bm{\sigma}^2$ of residual interference and noise of the $k^\mathrm{th}$ transmit antenna, i.e.,
\begin{equation}
p(\hat{s}_k\mid s_k)=\frac{1}{\pi \sigma^2_k}\cdot\mathrm{exp}\left(-{
\frac{\lvert\hat{s}_k-s_k \rvert^2}{\sigma^2_k}}\right).
\end{equation}
\section{Simulation Results}
We have simulated the turbo coded MIMO-OFDM system with the parameters presented in Table III. The simulation assumed a multipath fading PLC channel with 4 paths. Note that spacing between tones is set to be 2 KHz. When Doppler frequency shift happens barely in PLC system, the sub-carrier spacing of 2 KHz is harmless to system performance.
\begin{table}[htbp]
\caption{Simulation Parameter}
\begin{center}
\begin{tabular}{|l|l|}
\hline
\textbf{Parameters} &  \textbf{Value}\\
\hline
\hline
Transfer rate  &  2 Mbit/s\\
\hline
MIMO channel model& $3\times3$ quasi-static\\
\hline
Modulation   &  QPSK\\
\hline
Baseband Nyquist bandwidth  & 2 MHz \\
\hline
Number of OFDM sub-carriers & 1024\\
\hline
Spacing between tones  &  2 KHz\\
\hline
OFDM symbol duration & $500\mu$ sec\\
\hline
$\mathrm{PB}\_\mathrm{Size}$ of turbo Coding  &  264\\
\hline
Code rate&  1/2\\
\hline
AWCN parameter $A$ &  0.1\\
\hline
AWCN parameter $\Gamma$ &  0.01\\
\hline
\end{tabular}
\label{tab1}
\end{center}
\end{table}

Fig. 5 compares the BER of $3\times3$ MIMO-OFDM system with and without turbo coding. We set the frequency band to be 3.0 MHz-5.0 MHz. It is observed that turbo coding improves the system performance remarkably, with more than 2dB gain at BER=$10^{-3}$ and BER=$10^{-4}$ compared to the system without turbo coding. Thus the turbo coding method adopted by \textit{low voltage power line communication interoperability technical specification}\cite{b8} performs very well in MV PLC system.
\begin{figure}[htbp]
\centerline{\includegraphics[width=8cm,height=6cm]{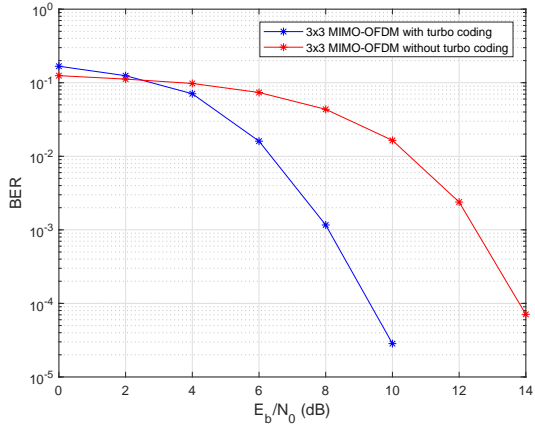}}
\caption{BER performance comparison of MIMO-OFDM with and without turbo coding }
\label{fig5}
\end{figure}
\begin{figure}[htbp]
\centerline{\includegraphics[width=8cm,height=6cm]{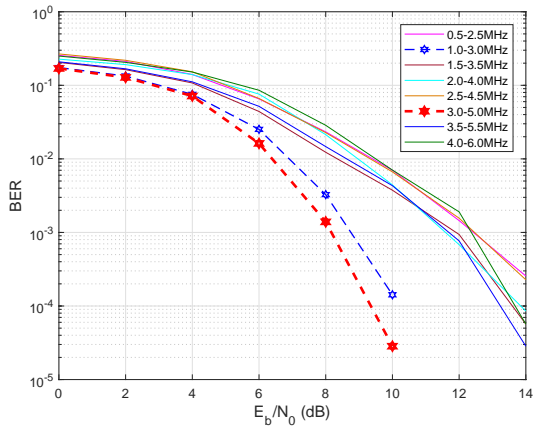}}
\caption{BER performance comparison of different frequency bands}
\label{fig6}
\end{figure}

The BER performance comparison of different frequency bands (with the same bandwidth of 2MHz) is given in Fig. 6. It can be observed that within the frequency range of 0.5 MHz-6.5 MHz, the best performance is achieved on frequency band of 3.0 MHz-5.0 MHz, whereas 1.0 MHz-3.0 MHz comes second. There is no significant difference between the other frequency bands. The frequency band of 3.0 MHz-5.0 MHz obtains at least 2dB gain at BER=$10^{-4}$ compared to the other frequency bands except for 1.0 MHz-3.0 MHz whose BER performance is a little inferior to 3.0 MHz-5.0 MHz with a gap of approximately 0.3dB between them at BER=$10^{-4}$. In practice, 1.0 MHz-3.0 MHz based PLC system suffers more serious noise pollution than 3.0 MHz-5.0 MHz. To sum up, 3.0 MHz-5.0 MHz is the optimal frequency band for 2 MHz MIMO-OFDM system in power line transmission.
\section{Conclusion}
With the access to the electric and dimension parameters of a section of MV underground power cable in Ganzhou city, we have modeled the $3\times 3$  MIMO channel in MV PLC. Based on this model, we presented a comprehensive communication link for MV power line communication. The turbo coding module and soft demodulation module were introduced specifically since the turbo coding method used in this paper was originally designed for PLC system in \textit{low voltage power line communication interoperability technical specification}\cite{b8}. The simulation results ensured that the turbo coding method performs well in MIMO-OFDM system for MV power line communication. Furthermore, by comparing the BER performance of different frequency bands, the optimal frequency band for 2 MHz PLC system was also investigated.
\section*{Acknowledgment}
This work was supported by State Grid Corporation of China. The measurement of channel modeling parameters of the selected section of MV power cable, including electric and dimension parameters are provided by State Grid Corporation of China.

\end{document}